\documentclass[fleqn,usenatbib,letters]{mnras}
\usepackage{newtxtext,newtxmath}

\usepackage[T1]{fontenc}

\usepackage{ulem} 

\usepackage{graphicx}	
\usepackage{amsmath}	
\usepackage{amssymb}	

\title{Correcting HIRES/Keck radial velocities for small systematic errors}

\author[L. Tal-Or et al.]{
Lev Tal-Or,$^{1,4}$\thanks{E-mail: levtalo@tauex.tau.ac.il}
Trifon Trifonov,$^{2}$
Shay Zucker$^{1}$
Tsevi Mazeh$^{3}$
and Mathias Zechmeister $^{4}$
\\
$^{1}$Department of Geophysics, Raymond and Beverly Sackler Faculty of Exact Sciences, Tel Aviv University, Tel Aviv, 6997801, Israel\\
$^{2}$Max-Planck-Institut f\"ur Astronomie, K\"onigstuhl 17, D-69117 Heidelberg, Germany\\
$^{3}$School of Physics and Astronomy, Raymond and Beverly Sackler Faculty of Exact Sciences, Tel Aviv University, Tel Aviv, 6997801, Israel\\
$^{4}$Institut f\"ur Astrophysik, Georg-August-Universit\"at, Friedrich-Hund-Platz 1, 37077 G\"ottingen, Germany
}

\date{Accepted 2018 November 28. Received 2018 November 28; in original form 2018 June 14}

\pubyear{2019}

\begin{document}
\label{firstpage}
\pagerange{\pageref{firstpage}--\pageref{lastpage}}
\maketitle

\begin{abstract}
The HIRES spectrograph, mounted on the $10$-m Keck-I telescope, belongs to a small group of radial-velocity (RV) instruments that produce stellar RVs with long-term precision down to $\sim1$\,m\,s$^{-1}$. In $2017$, the HIRES team published $64,480$ RVs of $1,699$ stars, collected between $1996$ and $2014$. In this bank of RVs, we identify a sample of RV-quiet stars, whose RV scatter is $<10$\,m\,s$^{-1}$, and use them to reveal two small but significant nightly zero-point effects: a discontinuous jump, caused by major modifications of the instrument in August $2004$, and a long-term drift. The size of the $2004$ jump is $1.5\pm0.1$\,m\,s$^{-1}$, and the slow zero-point variations have a typical magnitude of $\lesssim1$\,m\,s$^{-1}$. In addition, we find a small but significant correlation between stellar RVs and the time relative to local midnight, indicative of an average intra-night drift of $0.051\pm0.004$\,m\,s$^{-1}$\,hr$^{-1}$. We correct the $64,480$ HIRES RVs for the systematic effects we find, and make the corrected RVs publicly available. Our findings demonstrate the importance of observing RV-quiet stars, even in the era of simultaneously-calibrated RV spectrographs. We hope that the corrected HIRES RVs will facilitate the search for new planet candidates around the observed stars.
\end{abstract}


\section{Introduction}
\label{sec1}

The fact that stellar radial-velocity (RV) measurements are prone to instrumental systematic errors is known for more than a century \citep[e.g.,][]{Albrecht1914}. Over the years, it has become a common practice to observe standard RV stars to calibrate the instrumental zero-point RVs and enable inter-comparison between different instruments \citep*[see, e.g.,][for historic review]{Stefanik1999}. \citet*{Udry1999} reported on regularly following a set of $50$ standard stars with CORAVEL \citep{Baranne1979}, and using them to correct for instrumental zero-point drifts of $>1$\,km\,s$^{-1}$. In their pioneering work with the HF absorption cell as a simultaneous reference, \citet*{Campbell1988} achieved long-term precision of $\sim13$\,m\,s$^{-1}$ by correcting the stellar RVs for run-to-run zero-point variations, which they measured using the non-variable stars of their survey.

The technological progress in the past three decades led to significant improvements in RV measurement precision. The advent of simultaneous calibration methods and environmentally stabilized spectrographs, enabled instruments, such as HIRES \citep{Vogt1994HIRES} and HARPS \citep{mayor03}, to reach long-term precision of $\sim1$\,m\,s$^{-1}$. ESPRESSO now promises to break the $1$\,m\,s$^{-1}$ limit in the coming years \citep[e.g.,][]{Pepe2014}. However, these instruments heavily rely on sophisticated calibration schemes, and the practice of using standard RV stars to correct for systematic errors was largely abandoned.

HIRES is a general-purpose high-resolution visible-light slit spectrograph mounted on the $10$-m Keck-I telescope \citep{Vogt1994HIRES}. To enable precision RV measurements, an absorption Iodine cell was placed in front of the slit, which enabled measurement precision down to $\sim3$\,m\,s$^{-1}$ \citep{Butler1996,Vogt2000}. In August $2004$ a major upgrade of HIRES was performed, which improved the limiting RV precision by factor $\sim3$ \citep{Butler2006}. HIRES shares time on the Keck-I telescope with several other instruments. Therefore, observations of bright stars for precision RV measurements are typically scheduled around bright times. Over the years, HIRES was extensively used to search for exoplanets around bright dwarf stars \citep[e.g.,][]{Vogt2002,Vogt2005,Butler2003,Butler2004,Butler2006,Cumming2008}. Although the survey included many RV-stable stars, their RVs were used mainly to monitor the long-term stability of the instrument, but not to correct for systematic zero-point variations \citep[e.g.,][]{Vogt2002aspc,Vogt2015}.

In this paper we use publicly-available HIRES measurements to find and correct for small systematic RV variations. For the benefit of the exoplanet community, we upload the corrected RVs, and the code we used to calculate the correction, to dedicated websites. 

\section{The public HIRES RVs and auxiliary data}
\label{sec2}

\begin{figure}
\resizebox{\hsize}{!}
{\includegraphics{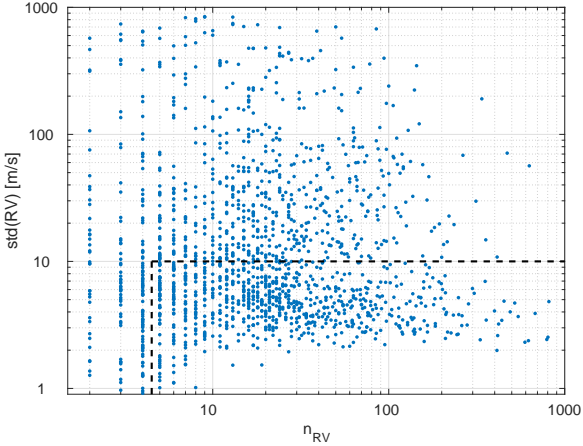}}
\caption{The HIRES targets published by \citet{Butler2017}: std(RV) and $n_{\rm RV}$ per star. We used only the stars with $n_{\rm RV}\geq5$ and std(RV) $<10$\,m\,s$^{-1}$ in the search of systematic effects (dashed line). The Y axis is limited to $1$\,km\,s$^{-1}$ as the more variable stars are irrelevant to the work presented here.
}
\label{fig1}
\end{figure}

In a legacy paper, \citet{Butler2017} published $64,480$ observations of $1,699$ stars\footnote{Available for download at \url{https://ebps.carnegiescience.edu/data/}}, collected with HIRES between $1996$ and $2014$ as part of a long-term search for exoplanets. The vast majority of the observed stars were chromospherically-inactive F, G, K, and M dwarfs, and most of them were observed over a time baseline of three years or more.

In order to search for systematic effects in the HIRES data, we analyzed the un-binned data provided by \citet{Butler2017}. Each observation is described by its Barycentric Julian Date (BJD), exposure time ($t_{\rm exp}$), median number of photons per pixel ($\overline{n_{\rm phot}}$), RV, RV uncertainty, and two indices of chromospheric activity.

Figure \ref{fig1} shows the RV standard deviation, std(RV), as a function of the number of RVs per star ($n_{\rm RV}$) in the HIRES data. For robustness against outliers and small-number statistics, we estimated std as $1.48$ times the median absolute deviation (MAD) around the median \citep[e.g.,][]{Rousseeuw1993}. The std(RV) distribution peaks at $5$--$6$\,m\,s$^{-1}$, with tails extending down to $\sim2$\,m\,s$^{-1}$ and up to $>100$\,m\,s$^{-1}$. For the purpose of studying systematic RV errors we focused only on reliable RV-quiet stars, which we defined as those with $n_{\rm RV}\geq5$ and std(RV) $<10$\,m\,s$^{-1}$. We found $797$ such stars, with a total of $39,930$ RVs.

\section{Small systematic errors in HIRES RVs}
\label{sec3}

\subsection{nightly zero point RV variations}

\begin{figure*}
\resizebox{\hsize}{!}
{\includegraphics{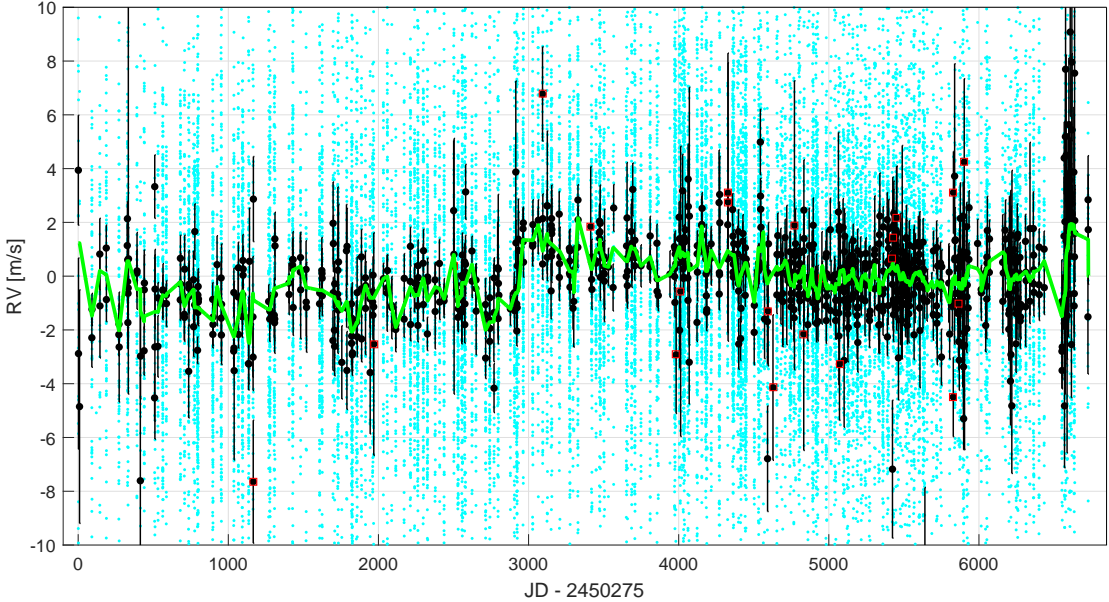}}
\caption{Systematic variations of the HIRES zero-point: individuals RVs of RV-quiet stars are shown in cyan, and the calculated NZPs are shown in black. NZPs that were derived from too few RVs ($<3$) are marked with red boxes. The green line shows our adopted NZP model: a moving weighted-average with a $50$-day window. The Y axis was limited to $\pm10$\,m\,s$^{-1}$ to enable inspecting the small-size NZP variations.}
\label{fig2}
\end{figure*}

\begin{figure}
{\includegraphics[width=90mm]{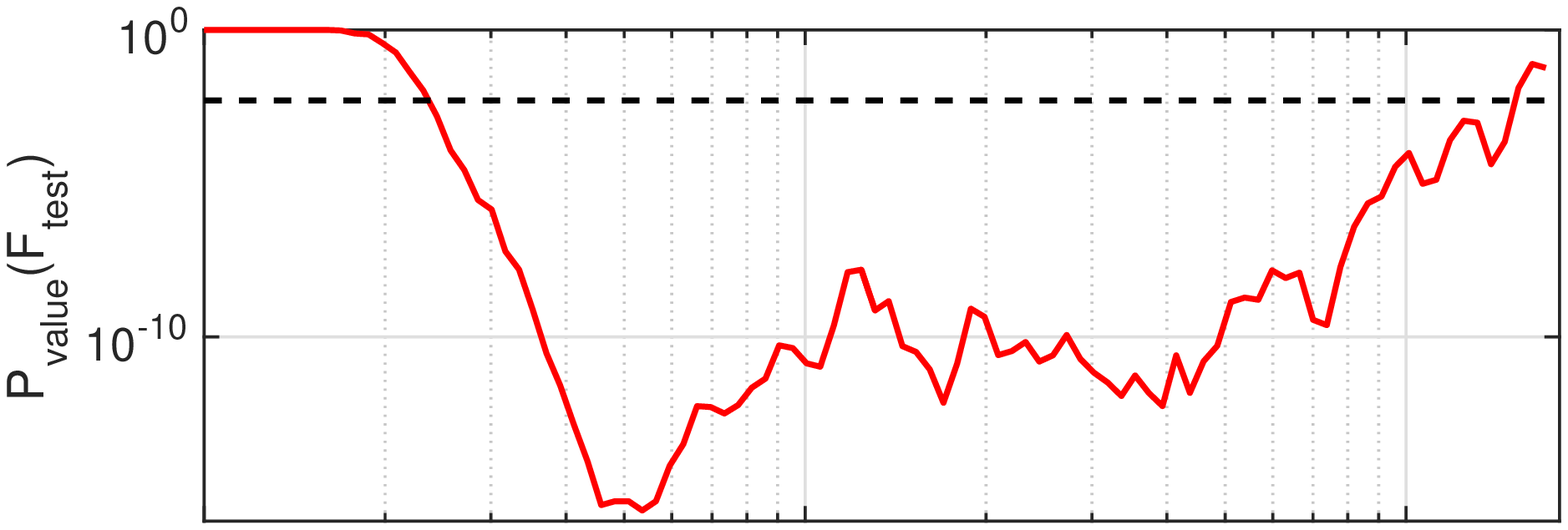}}
{\includegraphics[width=90mm]{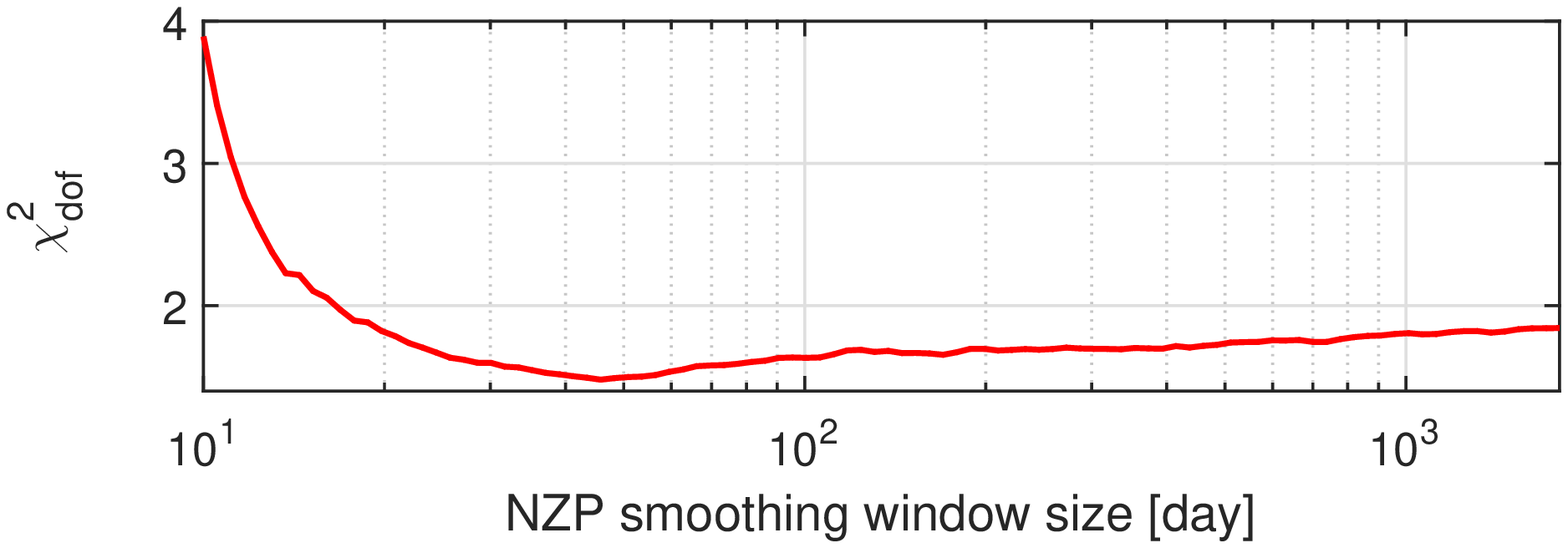}}
\caption{Searching for the typical time-scale of HIRES NZP variations: the upper and lower panels show the $p(F_{\rm test})$-value and $\chi^2_{\rm dof}$ statistics of subtracting a moving weighted average from the NZPs. The dashed line in the upper panel is the p-value $=0.005$ line, which is usually considered as the critical value for significant evidence \citep{Benjamin2018}.}
\label{fig3}
\end{figure}

We calculated an instrumental nightly zero-point RV (NZP) for each night in which at least three different RV-quiet stars were observed. The NZP of each observing night was taken as the weighted average RV of the RV-quiet stars that were observed in that night. In the $18.4$ years of HIRES data provided by \citet{Butler2017}, we found 
$913$ nights for which we could calculate a NZP. For completeness, we summarize our NZP calculation method in the Appendix.

Figure \ref{fig2} shows the RVs used to calculate the NZPs, and the derived NZPs. The NZPs have a weighted rms of $\sim1.20$\,m\,s$^{-1}$ and an effective uncertainty of $\left<\delta{\rm NZP}\right>_{-2} \sim 0.74$\,m\,s$^{-1}$, which gives a $\chi^2_{\rm red}$ of $\sim2.61$ \citep{Zechmeister2013}. Hence, the NZPs reveal an additional source of systematic RV scatter, on top of the internal RV uncertainties. We note that the minimum criteria for RVs to be used in calculating the NZPs were chosen after extensive optimization. More conservative criteria, such as $n_{\rm RV}\geq10$, std(RV) $<8$\,m\,s$^{-1}$, and $n_{\rm RV}/{\rm night}\geq5$, lead to very similar set of NZPs as presented in Fig. \ref{fig2}. Specifically, the typical difference between the NZPs calculated with the two different sets of criteria is twice as small as the typical NZP uncertainty.

Several features are present in HIRES NZPs: a discontinuous jump at JD $\sim2453225$ ($2004$ August $7$), a slow variation on time-scales of a few years, and an increased NZP scatter in $2014$, possibly showing an upward drift. The $2004$ jump is well explained by the major upgrade of HIRES \citep{Butler2017}. Comparing the three-year average zero point before and after the 2004 upgrade, we measured a jump of $1.5\pm0.1$\,m\,s$^{-1}$. The reasons behind the slow NZP variations, as well as the $2014$ variability, are not known to us.

In order to find the typical time-scale of the zero-point variations, we calculated a moving weighted-average of the NZPs, varying the window size from $10$ to $1800$ days. A moving weighted-average can be viewed as a non-parametric model of the data, with an effective number of parameters equal to the time-span of the data divided by the window size. Figure \ref{fig3} shows the $\chi^2_{\rm dof}$ and the $p(F_{\rm test})$-value of subtracting the moving average from the NZPs. Both statistics have a minimum at a smoothing window of $\sim50$ days. Since the nights allocated to HIRES were typically grouped around bright times, the minimum values indicate that the most prominent systematic effect is actually run-to-run zero-point variations, rather than night-to-night ones. This is similar to the effect detected by \citet{Campbell1988} in their survey.

We have thus adopted the $50$-day moving average of the NZPs as our model for the long-term systematic RV variations in HIRES. It is plotted in Fig. \ref{fig2} with a green solid line. The smoothing lead to a decrease of the effective NZP uncertainty from $\sim 0.74$\,m\,s$^{-1}$ to $\sim 0.29$\,m\,s$^{-1}$. In addition, the short smoothing window obviated the need to model the $2004$ jump separately, as it can be viewed as yet another run-to-run jump, with one difference: while virtually all other jumps go stochastically up and down, the $2004$ jump changed the instrument's zero point quasi permanently.

Using the adopted NZP model we corrected all HIRES RVs by subtracting the model from each RV according to its night of observation. To avoid self biasing, we calculated the NZP model for each RV-quiet star by using the RVs of all other RV-quiet stars. The model uncertainties were added in quadrature to the internal RV uncertainties. Since the adopted model averaged a few adjacent NZPs for each night, the mean model uncertainty is $\sim0.3$\,m\,s$^{-1}$, and higher than $1.0$\,m\,s$^{-1}$ in nine nights only.

The mean absolute value of the correction is $\sim0.6$\,m\,s$^{-1}$. Hence, the correction removed from the HIRES RVs a small but significant systematic effect. The correction lowered the median std(RV) of the RV-quiet stars from $\sim4.7$ to $\sim4.6$\,m\,s$^{-1}$. Since their median RV uncertainty ($\overline{\delta{\rm RV}}$) is $\sim1.5$\,m\,s$^{-1}$, we conclude that their RV scatter, even after the NZP correction, is still dominated by either intrinsic or additional systematic RV variations.

\subsection{intra-night RV drift}

\begin{figure}
\resizebox{\hsize}{!}
{\includegraphics{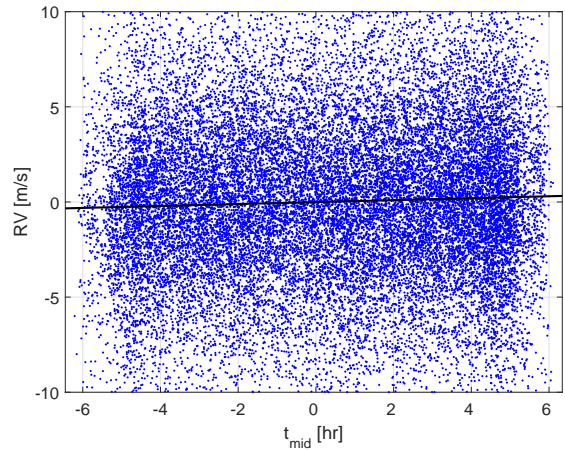}}
\caption{Systematic RV-$t_{\rm mid}$ correlation in HIRES: NZP-corrected RV-quiet star RVs are shown in blue, and the black line shows the adopted linear correction model, with a slope of $0.051\pm0.004$\,m\,s$^{-1}$\,hr$^{-1}$. The Y axis was limited to $\pm10$\,m\,s$^{-1}$ to enable inspecting the small-size effect.}
\label{fig4}
\end{figure}

We used the NZP-corrected RVs and the auxiliary data of the observations to look for additional systematic effects in HIRES RVs. Using the criteria of \citet{Benjamin2018}, who set the threshold for significant findings to p-value $=0.005$, 
we found one small but significant correlation: between the RVs and the time relative to local midnight ($t_{\rm mid}$). The linear RV-$t_{\rm mid}$ correlation, shown in Fig. \ref{fig4}, have a $p(F_{\rm test})$-value of $\sim3\cdot10^{-8}$, and a slope of $0.051\pm0.004$\,m\,s$^{-1}$\,hr$^{-1}$. For completeness, we further corrected the RVs for this small RV-t$_{\rm mid}$ correlation. 
We also checked whether a second or third-order polynomials would better describe the RV-$t_{\rm mid}$ relation, but found the improvement to be insignificant.

\section{The impact of the correction on periodic RV signals}
\label{sec4}

 \begin{table*}
 	\centering
 	\caption{Five representative lines from the online table of HIRES RVs corrected for small systematic errors.}
 	\label{tab1}
 	\begin{tabular}{lccccccccc} 
 		\hline
 		Target & BJD & RV [m\,s$^{-1}$] & $\delta$RV [m\,s$^{-1}$] & S-index & H-index & $\overline{n_{\rm phot}}$ & $t_{\rm exp}$ [sec] & CV [m\,s$^{-1}$] & $\delta$CV [m\,s$^{-1}$] \\
 		\hline
GJ\,3470 & $2456196.12775$ & $-1820.918$ & $2.681$ & $1.5725$ & $0.05486$ & $8749$ & $860$ & $0.178$ & $0.238$ \\
GJ\,3470 & $2456203.09433$ & $-1735.865$ & $2.713$ & $1.6339$ & $0.05594$ & $8610$ & $914$ & $0.545$ & $0.267$ \\
GJ\,3470 & $2456290.09917$ & $451.066$ & $3.450$ & $1.7408$ & $0.05560$ & $7128$ & $804$ & $0.244$ & $0.522$ \\
GJ\,3470 & $2456325.97211$ & $1258.111$ & $2.792$ & $1.7354$ & $0.05556$ & $8695$ & $1139$ & $0.339$ & $0.258$ \\
GJ\,3470 & $2456326.96973$ & $1266.413$ & $2.573$ & $1.5718$ & $0.05515$ & $8711$ & $846$ & $0.337$ & $0.258$ \\
 		\hline
 	\end{tabular}
 \end{table*}

The corrected HIRES RVs, as well as the correction values, are given in an online table\footnote{Table of the radial velocities is available at the CDS via anonymous ftp to \url{cdsarc.u-strasbg.fr} (\url{130.79.128.5}) or via \url{http://cdsarc.u-strasbg.fr/viz-bin/qcat?J/MNRAS/484/L8}}. The table has a similar format to the online version of table 1 of \citet{Butler2017}, with two values appended to each line: the correction value (CV) and its uncertainty ($\delta$CV). Table \ref{tab1} shows five representative lines from the online table.

Since almost all the stars with $n_{\rm RV}\geq10$ had std(RV) of $\gtrsim2$\,m\,s$^{-1}$ before the correction, and since the typical correction values are $\lesssim1$\,m\,s$^{-1}$, the correction usually had a very small effect on the RV scatter of individual stars. However, the correction does have some typical time-scales, and it is interesting to see how it impacts specific frequencies in some stellar RV power spectra.

\begin{figure}
\resizebox{\hsize}{!}
{\includegraphics{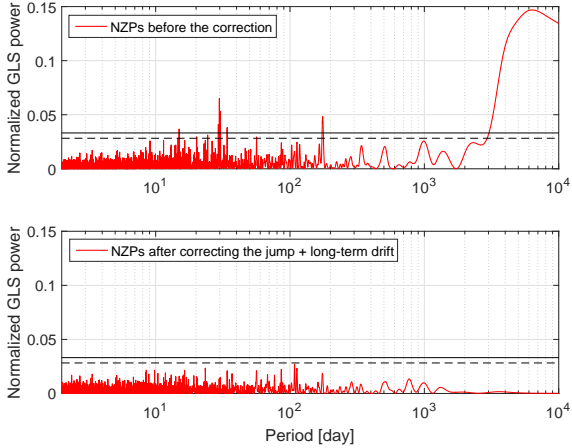}}
\caption{Normalized GLS periodograms of the NZPs, before and after subtracting from the NZPs the $2004$ jump and the long-term drift. The dashed and solid black lines mark the $0.001$ and $0.0001$ FAP lines.}
\label{fig5}
\end{figure}

The upper panel of Fig. \ref{fig5} shows the normalized GLS periodogram \citep{Zechmeister2009a} of the NZPs. The three highest peaks, in descending power, belong to periods of $\sim6,500$, $\sim30$, and $\sim175$ days. 
The lower panel of Fig. \ref{fig5} shows the same periodogram, after correcting the NZPs for the $2004$ jump and the long-term drift, by subtracting the $1.5$\,m\,s$^{-1}$ jump and a moving weighted-average filter of a $1,000$-day window. The correction removed all significant periodogram peaks, pushing the highest peak below a false-alarm probability (FAP) of $0.001$. This indicates that the $30$-day and the $175$-day peaks probably emerged from the window function, coupled with the jump and the long-term NZP variations. 
Similarly, we expect the adopted correction, detailed in Section \ref{sec3}, to impact mainly long-period low-amplitude RV signals, and spurious signals arising from coupling the systematic variability with the window function of each star.

\begin{figure}
\resizebox{\hsize}{!}
{\includegraphics{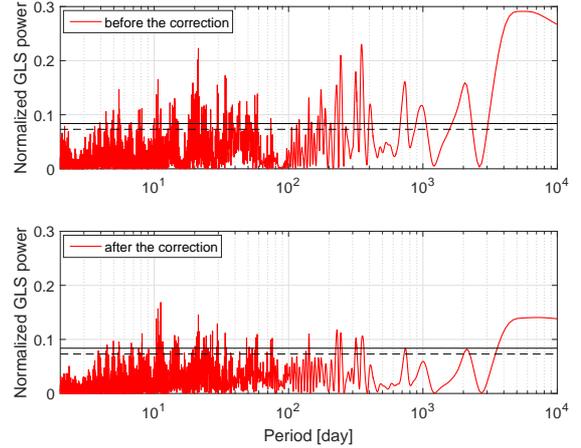}}
\caption{Normalized GLS periodograms of HIRES RVs of HD\,10476, before and after correcting for the systematic effects. The dashed and solid black lines mark the $0.001$ and $0.0001$ FAP lines.}
\label{fig6}
\end{figure}

In order to demonstrate such effect, Fig. \ref{fig6} shows the normalized GLS periodogram of the HIRES RVs of HD\,10476, before and after the correction. HD\,10476 appears in table 2 of \citet{Butler2017} as a planet candidate with $P\sim5,000$ days and $K\sim2$\,m\,s$^{-1}$. Not only the correction suppressed all periodogram peaks to higher FAP levels, it also made the $5,000$-day peak less significant than the shorter-period $11$-day one. We believe that a repeated search for candidates, which would use the corrected RVs and a similar method to the one applied by \citet{Butler2017}, might remove HD\,10476 from the candidates list.

\begin{figure}
\resizebox{\hsize}{!}
{\includegraphics{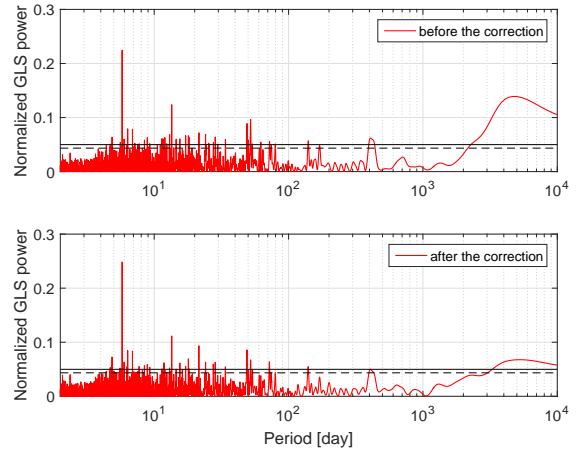}}
\caption{Same as Fig. \ref{fig6}, but for HD\,1461.}
\label{fig7}
\end{figure}

Figure \ref{fig7} shows the normalized GLS periodogram of the HIRES RVs of HD\,1461, which is known to host two short-period super-Earth planets, with orbital periods of $5.77$ days \citep{Rivera2010} and $13.5$ days \citep{Diaz2016}. In addition, the star is claimed to show a long-period $\sim4,000$-day signal, probably related to activity \citep{Diaz2016}. The correction suppressed the long-period signal to a FAP level higher than the signals of the two published planets. However, it also slightly suppressed the $13.5$-day signal, while the $5.77$-day signal became more significant. Therefore, it could be interesting to perform combined analysis of the corrected HIRES RVs and the HARPS RVs published by \citet{Diaz2016}.

\section{Summary and conclusions}
\label{sec5}

We have presented a correction of the HIRES RVs, published by \citet{Butler2017}, for two small but significant systematic effects: variations of the instrumental zero-point RV of $\lesssim1$\,m\,s$^{-1}$, and an intra-night RV drift of $\sim0.05$\,m\,s$^{-1}$\,hr$^{-1}$. Periodogram analysis of the NZPs, and of planet candidates from \citet{Butler2017}, shows that the correction affects mainly low-amplitude ($\lesssim2$\,m\,s$^{-1}$) and long-period ($\gtrsim2,000$ days) signals. We provide the exoplanet community with the corrected HIRES RVs, which are now more self consistent over the $\sim18$ years of observations. To facilitate verification or reproduction of our results, the code we used to calculate the corrections is also available for download\footnote{\url{https://github.com/levtalor/correcting-hires-Keck-data}}.

The presented results suggest that repeated observations of RV-quiet stars is important even in the era of simultaneously-calibrated and stabilized RV spectrographs, and that it can reveal systematic errors well below the noise level of the instrument. The methods applied in this work are simple and straight forward. They were used successfully in the past \citep[e.g.,][]{Campbell1988}, as well as recently \citep[e.g.,][]{Trifonov2018}. They can be applied to any precision RV survey, given that enough RV-quiet stars are observed every night.

Presently, more and more precision RV instruments become operational, and the precision is being pushed to the $\sim0.1$\,m\,s$^{-1}$ level \citep[e.g.,][]{Fischer2016}. In order to identify and correct for instrumental systematic RV errors, we urge the observatories to keep observing RV-quiet stars on a nightly basis. In addition, we encourage the different RV surveys to publish their full bank of RVs, as done by \citet{Butler2017}, for the whole exoplanet community to analyse.


\section*{Acknowledgements}

This research was supported by the ISRAEL SCIENCE FOUNDATION
(grant no. 848/16).



\bibliographystyle{mnras}
\bibliography{LevTalOr}



\appendix

\section{NZP calculation method}
\label{app1}

To avoid overweighting a single star in a given night, we first binned repeated observations of the same star in a given night by taking the weighted average of its RVs in that night, and its RV scatter in that night as its RV error. Then, we subtracted, from each-star RVs, its-own weighted-average RV, $\mu_k$, with the weights calculated from its internal RV uncertainties as $\delta{\rm RV}_k^{-2}$. The average-RV uncertainty was calculated via
\begin{equation}
\delta\mu_k = {\rm max}\{(\sum_k{\rm \delta{\rm RV}_{k}^{-2}})^{-0.5},{\rm wstd}_k \cdot n_{{\rm RV}}^{-0.5}\},
\label{eq1}
\end{equation}
where wstd$_k$ is the weighted std of its RVs. Then, $\delta\mu_k$, was added in quadrature to its RV uncertainties. This way we down-weighted the RVs of low-quality RV-quiet stars.

The NZP of each survey night (NZP$_n$) was taken as the weighted-average RV of the RV-quiet stars observed that night, after removing $10\sigma_k$ outliers per star and $5\sigma_n$ outliers per night, as long as they deviated by $>10$\,m\,s$^{-1}$ from the mean. To avoid underestimating $\delta{\rm NZP}_n$, regardless whether the RV scatter of that night was dominated by intrinsic scatter or by the $\delta{\rm RV}$s of the observed stars,
we took 
\begin{equation}
\delta{\rm NZP}_n = {\rm max}\{(\sum_k{\rm \delta{\rm RV}_{k,n}^{-2}})^{-0.5},{\rm wstd}_n \cdot n_{{\rm RV},n}^{-0.5}\},
\label{eq2}
\end{equation}
where $n_{{\rm RV},n}$ is the number of RVs used to calculate the NZP of night $n$, and wstd$_n$ is the weighted std of the RVs in that night.

We note that the boundary between RV-quiet and RV-loud stars can be shifted to a value lower than $10$\,m\,s$^{-1}$, at the cost of reducing the number of stars used for the calculations. In addition, RV-quiet stars with true orbital or activity-induced signals were not excluded from NZP calculations. Replacing their RVs with their RV residuals, after subtracting a model for the true variation, is a possible improvement we consider for our algorithm.

Another possible improvement of the algorithm is to model $\mu_k$ and the NZPs simultaneously by maximising the likelihood
\begin{equation}
\ln L = \sum_{k,n}\frac{{\rm RV}_{k,n}-\mu_k-{\rm NZP}_n}{\delta{\rm RV}_{k,n}^2+\sigma_k^2}, 
\label{eq3}
\end{equation} 
where $\sigma_k^2$ is the so-called RV jitter, which should take into account unknown low-amplitude variations due to stellar activity or orbit. Such approach has the advantage of a self-consistent weighting of each-star RVs. 
However, it is computationally heavy, and estimating $\delta{\rm NZP}_n$ is not straight forward.

\bsp	
\label{lastpage}
\end{document}